# Precise calibration of Mg concentration in $Mg_xZn_{1-x}O$ thin films grown on ZnO substrates

Y. Kozuka[1,a], J. Falson[1], Y. Segawa[2], T. Makino[2], A. Tsukazaki[1], and M. Kawasaki[1,2]

[1]*Department of Applied Physics and Quantum-Phase Electronics Center (QPEC), University of Tokyo, Tokyo 113-8656, Japan*

[2]*Cross-Correlated Materials Research Group (CMRG) and Correlated Electron Research Group (CERG), RIKEN-Advanced Science Institute, Wako 351-0198, Japan*

[a]Electronic mail: kozuka@ap.t.u-tokyo.ac.jp

Abstract

The growth techniques for $Mg_xZn_{1-x}O$ thin films have advanced at a rapid pace in recent years, enabling the application of this material to a wide range of optical and electrical applications. In designing structures and optimizing device performances, it is crucial that the Mg content of the alloy be controllable and precisely determined. In this study, we have established laboratory-based methods to determine the Mg content of $Mg_xZn_{1-x}O$ thin films grown on ZnO substrates, ranging from the solubility limit of $x \sim 0.4$ to the dilute limit of $x < 0.01$. For the absolute determination of Mg content, Rutherford backscattering spectroscopy is used for the high Mg region above $x = 0.14$, while secondary ion mass spectroscopy is employed to quantify low Mg content. As a lab-based method to determine the Mg content, $c$-axis length is measured by X-ray diffraction and is well associated with Mg content. The interpolation enables the determination of Mg content to $x = 0.023$, where the peak from the ZnO substrate overlaps the $Mg_xZn_{1-x}O$ peak in standard laboratory equipment, and thus quantitative determination. At dilute Mg contents below $x = 0.023$, the localized exciton peak energy of the $Mg_xZn_{1-x}O$ films as measured by photoluminescence is found to show a linear Mg content dependence, which is well resolved from the free exciton peak of ZnO substrate down to $x = 0.0043$. Our results demonstrate that X-ray diffraction and photoluminescence in combination are appropriate methods to determine Mg content in a wide Mg range from $x = 0.004$ to $0.40$ in a laboratory environment.

## I. INTRODUCTION

Epitaxial oxide thin films and their interfaces have been the subject of intensive research for decades in the exploration of material properties beyond conventional semiconductors.[1] Among the oxides, ZnO is one of the most promising materials for versatile photonic and electronic applications. For instance, high-intensity ultraviolet light emitting diodes in ZnO-based *p-n* junctions have been realized in this material owing to the rapid development of ZnO thin film growth technique.[2] In the same stream, *p-n* junctions combined with a microcavity explore the possibility of realizing room-temperature exciton-polariton lasing, where coherent light is expected to emit from exciton-polariton Bose-Einstein condensates under extremely low injection current.[3] In a similar heterostructure, a high-mobility two-dimensional electron gas (2DEG) has been found to form at the interface of $Mg_xZn_{1-x}O$ and ZnO. Since such a 2DEG is accumulated due to the polarization mismatch between $Mg_xZn_{1-x}O$ and ZnO and without the aid of intentional doping,[4] the electron mobility is extremely high, exceeding 700 000 $cm^2$ $V^{-1}$ $s^{-1}$ at $x \approx 0.01$, which is comparable to the mobility of other well-known two-dimensional systems, such as AlGaAs/GaAs and SiGe/Si.[5]

These functional physical properties of ZnO are as a result of its large band gap (3.37 eV) and exciton binding energy (60 meV), as well as spontaneous polarization along *c*-axis due to inversion-asymmetric Wurtzite structure.[6] Importantly, these physical parameters may be tuned by the substitution of Zn ions in the host lattice by isovalent Mg ions, to form the ternary alloy of $Mg_xZn_{1-x}O$.[7] In a practical device utilizing these physical properties of ZnO, $Mg_xZn_{1-x}O$ thin films should be pseudomorphically grown on single-crystal ZnO substrates to maintain high crystalline quality. In such a system, it is crucial that the extent of Mg in the alloy be controllable so that practical device design may be achieved. However, at present such systematic work on the methods to determine Mg composition $x$ in $Mg_xZn_{1-x}O$ thin films grown on ZnO substrates have not been well established.

In this study, we have investigated various methods to determine the Mg composition in $Mg_xZn_{1-x}O$ thin films grown on ZnO substrates. In such a heterostructure, it is not possible to use methods that are unable to separate the signals from the film and the substrate, such as inductively coupled plasma atomic emission spectroscopy and electron probe microanalysis. In order to determine $x$, therefore, we employ Rutherford backscattering spectroscopy (RBS) for high Mg concentration, and secondary ion mass spectroscopy (SIMS) for those with low Mg concentration. Analytically quantified $x$ values in the $Mg_xZn_{1-x}O$ films are then used as the basis for calibration curves for the values of $c$-axis length deduced by X-ray diffraction (XRD)[8] and localized exciton (LE) emission energy revealed by photoluminescence (PL).[9] With using an almost linear relationship of these values with $x$, one can determine Mg content $x$ in $Mg_xZn_{1-x}O$ thin films on ZnO substrate with using standard laboratory techniques such as XRD and PL.

**II. EXPERIMENTAL**

Samples used in this study are listed in Table I with the Mg concentration determined as discussed below. The $Mg_xZn_{1-x}O$ thin films were grown on Zn-polar ZnO substrates (Tokyo Denpa Co.)[10] by molecular beam epitaxy at a substrate temperature of 750 °C using 7N Zn and 6N Mg sources. Distilled pure ozone (Meidensya Co.) was utilized as the oxygen source due to its extremely low impurity level.[5] The typical film thickness was 200 – 700 nm. The XRD measurement was carried out with a lab-based X-ray source with four-bounce Ge (220) monochrometer (X'Pert MRD, Panalytical Co. and SmartLab, Rigaku Co.). For PL measurement, we used a He-Cd laser (325 nm) for $x \leq 0.10$ and a Nd:YVO$_4$ laser (266 nm) for $x \geq 0.14$, as a result of the shifting band-gap energy of the $Mg_xZn_{1-x}O$ layer. The light intensity was ~ 20 mW/cm$^2$ at the sample surface.

**III. RESULTS AND DISCUSSIONS**

**A. Surface morphology**

In order to eliminate the possibility of structural defects affecting lattice constant such as granular structure, the surface morphology was examined by an atomic force microscopy as shown in Figs. 1(a) – 1(c) for the representative samples. The surfaces exhibit a step-and-terrace structure with the root-mean-square roughness of ~ 0.1 nm, regardless of Mg content investigated, which ensures the variation of the lattice constant purely originates from the alloying of MgO in ZnO.

**B. High Mg concentration**

First, RBS was utilized in determining the Mg content of the four most concentrated samples of this study. This technique is known to provide the absolute concentration without the need of calibration by standard samples. Figure 2 shows an example of the RBS spectra for a 200 nm-thick $Mg_xZn_{1-x}O$ film grown on ZnO substrate with the best-fit simulation curve of $x = 0.40$. The insets indicate the magnified spectra for Zn and Mg signals from $Mg_xZn_{1-x}O$ layer together with the simulation curves of $x = 0.4 \pm 0.04$ and $x = 0.4 \pm 0.1$. It is apparent from the data that a smaller fitting error is achieved when $x$ is estimated from Zn signal ($\Delta x = \pm 0.01$) than from Mg signal ($\Delta x = \pm 0.04$). Below, we refer to this $x$ value estimated from RBS as $x_{RBS}$.

Knowing this absolute value of $x_{RBS}$, we then employed laboratory-based XRD, to correlate the measured $c$-axis length of the $Mg_xZn_{1-x}O$ layer with the analytically determined Mg content.

Figure 3 shows $\theta$-$2\theta$ diffraction patterns around the ZnO (0004) peak. For the high Mg region, a second peak, corresponding to the $Mg_xZn_{1-x}O$ layer is clearly observed together with Laue fringes which reflect the thickness of the film. The $c$-axis length difference ($\Delta c$) between the ZnO substrate ($c$ = 520.4 pm) and the $Mg_xZn_{1-x}O$ thin film is plotted as a function of $x_{RBS}$ in Fig. 4 for the four samples. The best-fit line was found to be $\Delta c$ (pm) = $-0.069 \times x$. This relation is significantly different from that for relaxed $Mg_xZn_{1-x}O$ films grown on $Al_2O_3$ substrates,[7] as a result of the films being under epitaxial strain for this study; the in-plane lattice is coherently connected with that of ZnO substrate and is extended in comparison with the strain-free state. Here, we note that the present results also deviate from those reported by Nishimoto et al,[8] where $Mg_xZn_{1-x}O$ thin films are pseudomorphically grown on ZnO (0001) substrates as in the present study. In the previous work, Auger electron spectroscopy was used to probe the absolute value and depth profile of the Mg content of films. However, the calibration curve used for such quantification was originally formulated from $Mg_xZn_{1-x}O$ films grown on $Al_2O_3$ substrates, and hence brings into question possible errors in the original calibration due to differences in the sticking coefficient of Mg for films grown on ZnO as opposed to $Al_2O_3$.[11] We speculate that this gave significant error in the previous data and we insist to revise the relation by the data given in this paper. By using the relation shown in Fig. 4, interpolation gives estimates of Mg concentration, which we refer to as $x_{XRD}$, down to $x_{XRD}$ = 0.023. This value is as a result of the resolution limit of a regular lab-based monochrometer in XRD equipment, where ultimately the peak of the $Mg_xZn_{1-x}O$ and ZnO film cannot be separated, as displayed in Fig. 3, for $x$ = 0.011 (determined by SIMS as explained below). This limit may also be affected by the thickness or the quality of the film, which broaden the diffraction peaks.

**C. Low Mg concentration**

Although we have thus demonstrated that the Mg content can be determined from XRD, this is not applicable below $x \approx 0.02$ as the XRD peak from $Mg_xZn_{1-x}O$ layer is not clearly resolved from the peak of the ZnO substrate. RBS is not also applicable for the absolute determination of Mg concentration in this range because of the large error of $x \sim 0.01$. As another physical parameter which is dependent on Mg content, we focused on the exciton energy observed by PL, which is conventionally used to determine Al composition in (Al,Ga)As thin films grown on GaAs substrate.[12] For absolute calibration of the Mg content, we utilized SIMS measurements that were calibrated with a Mg ion-implemented standard sample. A series of depth profiles of the SIMS spectra are shown in Fig. 5(a), with the depth normalized by the thicknesses of the films. Due to finite inhomogeneity of Mg concentration along the depth, the peak of the histogram in Mg concentration is taken as a representative value as shown in Fig. 5(b).

Knowing the absolute Mg content of the dilute films, the energy dependence of the localized exciton (LE) luminescence on Mg concentration was investigated by PL for all samples at 100 K and 10 K as shown in Figs. 6(a) and 6(b), respectively. In the PL spectra, a peak corresponding to free excitons (FE) is clearly visible in ZnO ($x = 0$) at $T = 100$ K, while only weak intensity from FE was observed at $T = 10$ K as indicated by the asterisks. The intense peaks at lower energies originate from bound excitons, who were assigned observing the temperature dependence of such peaks (not shown). In the case of $Mg_xZn_{1-x}O$ films, the LE peak (indicated by asterisks in Fig. 6) appeared in addition to the FE peak from the ZnO substrate. At 10 K, additional broad peaks appeared at lower energies than the LE peaks, the origin of which is not clear at present because of its nonsystematic dependence on Mg content.

The energy difference ($\Delta E$) between LE emission energy from $Mg_xZn_{1-x}O$ layers and FE

emission energy from ZnO as a function of $x$ is plotted in Fig. 7(a) on a log-log scale. The overall feature indicates that $\Delta E$ has a quite good linear dependence on Mg content. In order to see the applicable range of the linear fitting, the relation is plotted in Fig. 7(b) on a linear scale, which indicates a nontrivial deviation of the LE exciton energies toward lower energy for high Mg content films from the extrapolated fitting line. This tendency is interpreted as stronger localization with higher Mg concentration. Thus, $c$-axis length from XRD is more appropriate to estimate Mg content at high Mg region due to its relative insensitivity to localized Mg concentration fluctuations compared to that of PL peak energies. Therefore, we provide a fitted relation of $\Delta E$ (eV) = 2.2 × $x$, only valid for that of $x \leq 0.023$. This LE energy dependence on Mg content is similar to previous results of $Mg_xZn_{1-x}O$ films grown on $Al_2O_3$ substrate.[9] The error bars of $x$ in Fig. 7 reflect the inhomogeneity of Mg concentration shown in the histogram of Fig. 5(b), which is negligible in our discussion.

## IV. CONCLUSION

In summary, we have established comprehensive means to determine the Mg content in $Mg_xZn_{1-x}O$ films grown on ZnO substrates by using standard lab-based XRD and PL techniques. For the high Mg content region of $x \geq 0.14$, the $c$-axis length is estimated by XRD and is well associated with $x$, where $x$ is calibrated by RBS. On the other hand, for the dilute region, a linear dependence between $x$ and LE exciton energy was obtained below $x \leq 0.015$, where $x$ is calibrated by SIMS. Extrapolation of $x$ using these two methods gives smooth connection of $x$ values in the range of $0.03 \leq x \leq 0.10$. This result can be widely used to determine Mg content in $Mg_xZn_{1-x}O$ films on ZnO substrates using XRD and PL in combination, depending on the Mg concentration range, and may form the infrastructure for continued research for the application of this promising material.

## ACKNOWLEDGEMENTS

This work was partly supported by "Funding Program for World-Leading Innovative R&D on Science and Technology (FIRST)" Program from the Japan Society for the Promotion of Science (JSPS) initiated by the Council for Science and Technology Policy.

Bertness, J. Appl. Phys. **93**, 3747 (2003).

**TABLE I.** $Mg_xZn_{1-x}O/ZnO$ samples used in this study. The $x$ values and their measurement methods used to determine $x$ are shown. $x$ values in brackets have not been analytically quantified, but interpolated through the X-ray diffraction calibration curve formulated.

| Sample No. | $x$ in $Mg_xZn_{1-x}O$ | Methods to determine Mg concentration | |
|---|---|---|---|
| | | Absolute value | Calibration |
| 79 | 0.0042 | SIMS | PL |
| 66 | 0.0056 | SIMS | PL |
| 77 | 0.0073 | SIMS | PL |
| 74 | 0.010 | SIMS | PL |
| 76 | 0.011 | SIMS | PL |
| 81 | 0.015 | SIMS | PL |
| 28 | (0.023) | | XRD, PL |
| 23 | (0.044) | | XRD, PL |
| 22 | (0.056) | | XRD, PL |
| 14 | (0.069) | | XRD, PL |
| 12 | (0.090) | | XRD, PL |
| 11 | (0.10) | | XRD, PL |
| 153 | 0.14 | RBS | XRD, PL |
| 147 | 0.20 | RBS | XRD, PL |
| 152 | 0.27 | RBS | XRD, PL |
| 168 | 0.40 | RBS | XRD, PL |

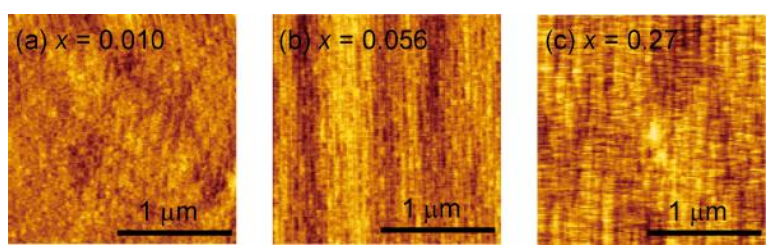

FIG. 1. Y. Kozuka *et al.*

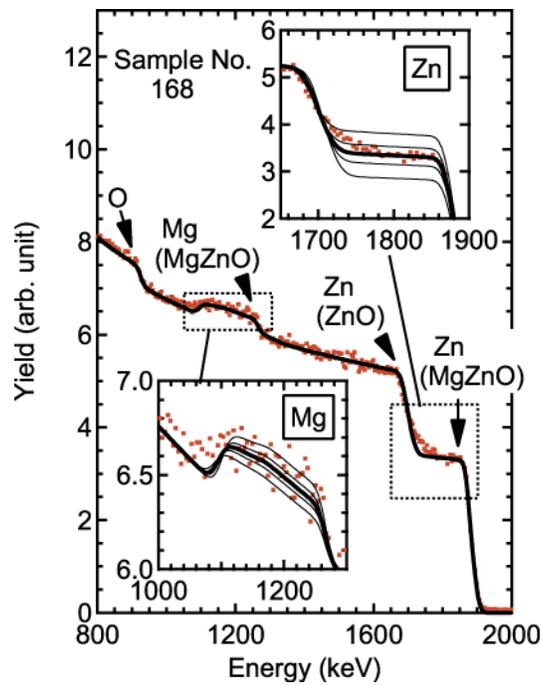

FIG. 2. Y. Kozuka *et al.*

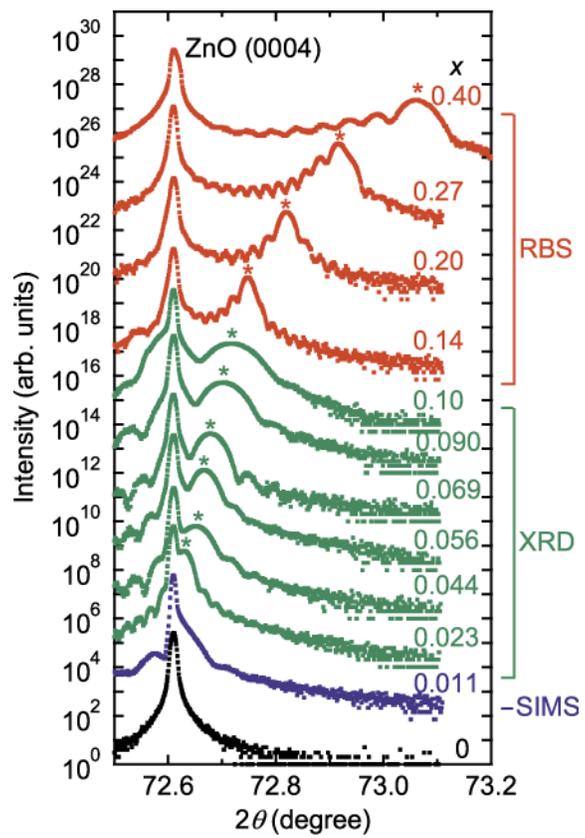

FIG. 3. Y. Kozuka *et al.*

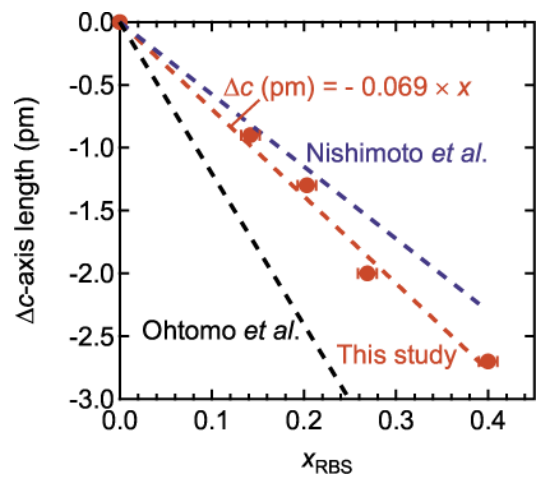

FIG. 4. Y. Kozuka *et al.*

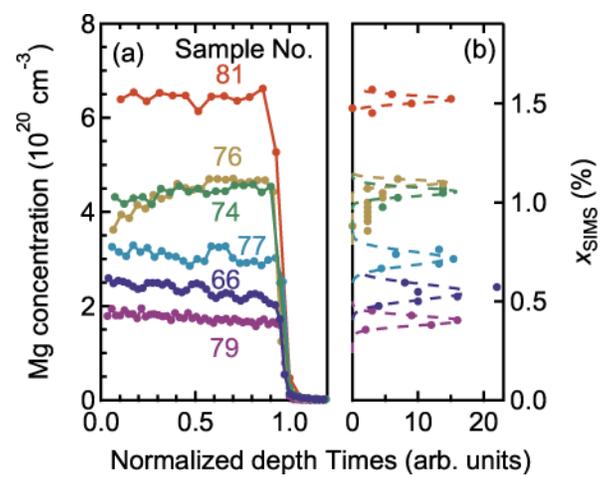

FIG. 5. Y. Kozuka *et al.*

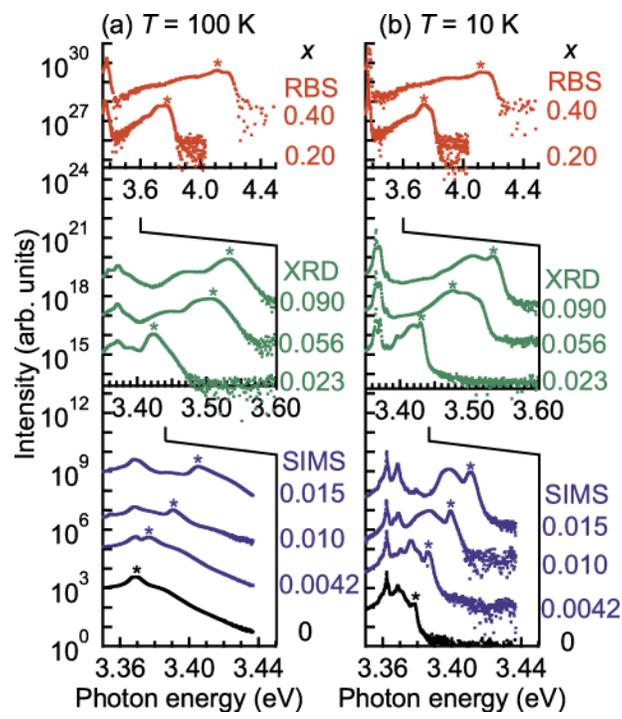

FIG. 6. Y. Kozuka *et al*.

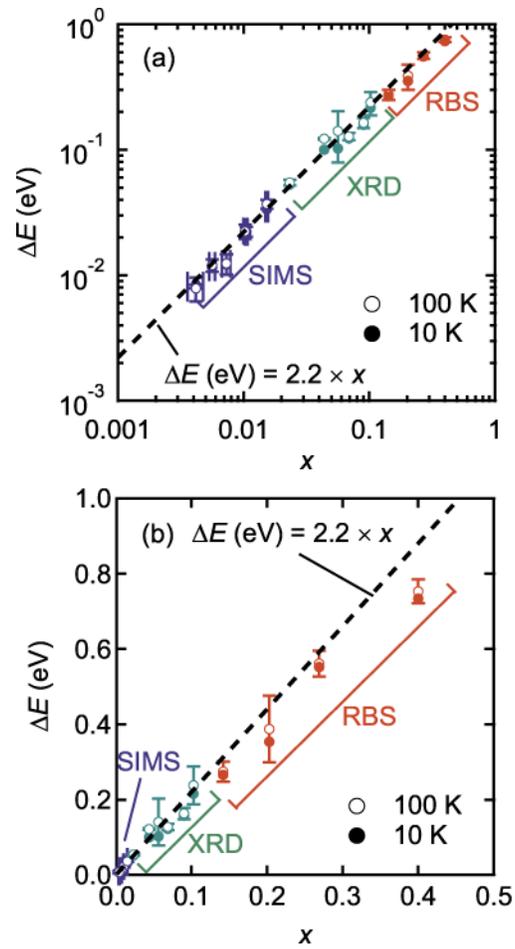

FIG. 7. Y. Kozuka *et al.*

**Figure Captions:**

FIG 1. Atomic force microscopy images of the surface morphology of $Mg_xZn_{1-x}O$ films for (a) $x = 0.010$, (b) $x = 0.056$, and (c) $x = 0.27$.

FIG 2. Rutherford backscattering spectrum for a 200 nm-thick $Mg_xZn_{1-x}O$ film on ZnO substrate. Dots are experimental data and the solid curves indicate the simulation. The onsets of the signals from Zn and Mg, and O are also indicated by arrows with the indication of the host layers in the brackets. Top and bottom insets are the magnifications around Zn and Mg signals, respectively, from the $Mg_xZn_{1-x}O$ layer together with the best-fit simulation curves, for $x = 0.40$ (bold), $x = 0.4 \pm 0.04$, and $x = 0.4 \pm 0.1$.

FIG 3. $\theta$-$2\theta$ X-ray diffraction around ZnO (0004) peak. The asterisks indicate the peaks corresponding to $Mg_xZn_{1-x}O$ layers. The methods to determine $x$ depend on Mg concentration ranges as indicated.

FIG 4. $c$-axis length difference ($\Delta c$) between the ZnO substrate and the $Mg_xZn_{1-x}O$ layer estimated from XRD as a function of Mg content determined from RBS. Previous results using ZnO substrate (Nishimoto *et al.*)[8] and $Al_2O_3$ substrate (Ohtomo *et al.*)[7] are also shown for comparison.

FIG 5. (a) Depth profile of Mg content measured by SIMS and (b) its histogram for low Mg samples. The dashed curves in (b) are the Gaussian fits to the histogram data.

FIG 6. Photoluminescence spectra measured (a) at 100 K and (b) at 10 K for representative $Mg_xZn_{1-x}O$ thin films. The asterisks correspond to emission peaks from FE for undoped ZnO and these from LE for $Mg_xZn_{1-x}O$ layers. The method to determine $x$ is also indicated (see Table I).

FIG 7. (a) Log-log and (b) linear plots of the energy difference ($\Delta E$) between LE emission from $Mg_xZn_{1-x}O$ films and FE emission from ZnO as a function of $x$ at 100 K and 10 K. The FE energy of ZnO is 3.377 eV at 10 K and 3.368 eV at 100 K. The methods to determine $x$ are also indicated (see Table I). The dashed lines are the fitting for the data below $x = 0.015$ at 100 K. The error bars in $x$ indicate the full width at half maximum in the histogram of Fig. 5(b), while those in $\Delta E$ are the full width at half maximum of LE peaks at 100 K.